\documentclass[reprint, aps,superscriptaddress,citeautoscript,floatfix]{revtex4-2}

\usepackage{graphics}
\usepackage{graphicx}
\usepackage{subfigure}
\usepackage{mathrsfs}
\usepackage{amsmath}
\usepackage{amssymb}
\usepackage{bbding}
\usepackage{makecell}
\usepackage{bm}
\usepackage[table,xcdraw]{xcolor}
\usepackage{color}
\usepackage{natbib}
\usepackage{bookmark}

\begin{document}
\date{\today}

\title{Effect of viscoelastic fluid on the lift force in lubricated contacts}

\author{Shiyuan Hu}
\email[]{shiyuan.hu@itp.ac.cn}
\affiliation{CAS Key Laboratory of Theoretical Physics, Institute of Theoretical Physics, Chinese Academy of Sciences, Beijing 100190, China}
\author{Fanlong Meng}
\affiliation{CAS Key Laboratory of Theoretical Physics, Institute of Theoretical Physics, Chinese Academy of Sciences, Beijing 100190, China}
\affiliation{School of Physical Sciences, University of Chinese Academy of Sciences, 19A Yuquan Road, Beijing 100049, China}
\affiliation{Wenzhou Institute, University of Chinese Academy of Sciences, Wenzhou, Zhejiang 325000, China}
\author{Masao Doi}
\email[]{doi.masao@a.mbox.nagoya-u.ac.jp}
\affiliation{Wenzhou Institute, University of Chinese Academy of Sciences, Wenzhou, Zhejiang 325000, China}
\affiliation{Oujiang Laboratory, Wenzhou, Zhejiang 325000, People’s Republic of China}

\begin{abstract}
We consider a cylinder immersed in viscous fluid moving near a flat substrate covered by an incompressible viscoelastic fluid layer, and study the effect of the fluid viscoelasticity on the lift force exerted on the cylinder. The lift force is zero when the viscoelastic layer is not deformed, but becomes non-zero when it is deformed. We calculate the lift force by considering both the tangential stress and the normal stress applied at the surface of the viscoelastic layer. Our analysis indicates that as the layer changes from the elastic limit to the viscous limit, the lift force decreases with the decrease of the Deborah number (De). For small De, the effect of the layer elasticity is taken over by the surface tension and the lift force can become negative. We also show that the tangential stress and the interface slip velocity (the surface velocity relative to the substrate), which have been ignored in the previous analysis, give important contributions to the lift force. Especially for thin elastic layer, they give dominant contributions to the lift force.

\end{abstract}

\maketitle

\section{Introduction}
Particles interacting with deformable interfaces immersed in a viscous fluid are common. Examples include lubricated elastic contacts in tribology~\cite{Hamrock04}, cartilaginous joints~\cite{Hou92}, and transport of red blood cells and vesicles~\cite{Abkarian02, Grandchamp13}. In such contacts, the two surfaces are separated with a finite gap filled with fluid. The sliding friction is determined by the viscous shear stress of the fluid, and is essentially determined by the gap distance between the surfaces. To prevent direct contact between the surfaces, one needs to have a force that tends to increase the distance between the sliding surfaces. Such force is called the lift force.

Reynolds~\cite{Reynolds86} first addressed to this problem and considered the situation that a cylinder is moving parallel to a flat substrate in a Newtonian fluid. He showed that the lift force is zero in this case. The absence of lift force for symmetric gap is also known as a consequence of the so-called scallop theorem at low Reynolds number~\cite{Purcell77}. To generate a nonzero lift, mechanisms that break the time-reversal symmetry are required~\cite{Bureau23}. 

The lift force is important in lubrication and has been extensively studied in hydrodynamics and mechanical engineering. Recently, the lift force in soft materials has attracted much attention because of its relevance in bio-medical applications. The lift force of rigid objects covered by soft elastic materials has been calculated~\cite{Sekimoto93, Skotheim04, Skotheim05}. Various types of deformable substrates have been studied, including polymer brushes~\cite{Sekimoto93, Davies18}, elastic membranes~\cite{Rallabandi18, Daddi18}, compressible and incompressible elastic solids~\cite{Skotheim04, Skotheim05, Urzay07, salez15, Saintyves16, Rallabandi17, Zhang20, Kargar21, Bertin22}, and viscoelastic solids~\cite{Pandey16, Kargar21}. 

Calculation of the lift force requires solving the hydrodynamic equation for the gap fluid coupled with the constitutive response of the substrate; for small deformations there are two prevailing approaches, one based on a perturbation expansion of the nonlinear Reynolds equation, and the other based on the Lorentz reciprocal theorem~\cite{Masoud19}. 

In the previous studies, attention has been focused on how the layer is deformed by the fluid pressure acting normal to the surface, and the tangential shear stress has been ignored. Such treatment may be justified for compressible layer, but not for incompressible layer where the surface can only be deformed via shear. The effect of tangential shear stress can be important~\cite{Chandler20}, especially for incompressible fluid-like layer. 

In this work, we study the effect of viscoelastic fluid layer covering the substrate on the lift force taking into account of both the tangential and the normal stresses acting on the layer surface. The present work is motivated by the mucus layer, which is a fluid-like material covering the luminal surfaces in humans and most animals and serving a variety of functions~\cite{Lai09, Atanasova19}. In the respiratory system the mucus overlays with a thinner periciliary layer, forming the airway surface liquid of thickness about 10 $\mu$m. The surface liquid is responsible for the capture and removal of foreign particles of varying sizes from micrometer to submillimeter scale~\cite{Wang05}. 

In the present study, we regard the mucus layer as a Maxwell fluid and conduct hydrodynamic calculation. We assume that the fluid has a single relaxation time $\tau$ and calculate the lift force with varying the Deborah number (De), the ratio of the relaxation time to the characteristic deformation time. We solve the displacement field of the layer considering the continuity conditions of the tangential and normal stresses at the layer surface. Our system reduces to the system of elastic layer for large De and to viscous fluid layer for small De. We show that for a thin layer the tangential stress and the nonuniform slip velocity (the velocity of layer surface relative to the substrate surface) play important roles in the lift force. In the elastic layer limit, our results give corrections to previous analysis. In the viscous layer limit, where the lift force is governed by the surface tension between the layer fluid and the outer fluid, we find that the lift force can become negative. This result is further confirmed by directly solving the Stokesian hydrodynamics in the lubrication limit. 

\section{Theoretical modeling}\label{sec_2}
\begin{figure}[b]
\centering
\includegraphics[bb=25 15 365 240, scale=0.55,draft=false]{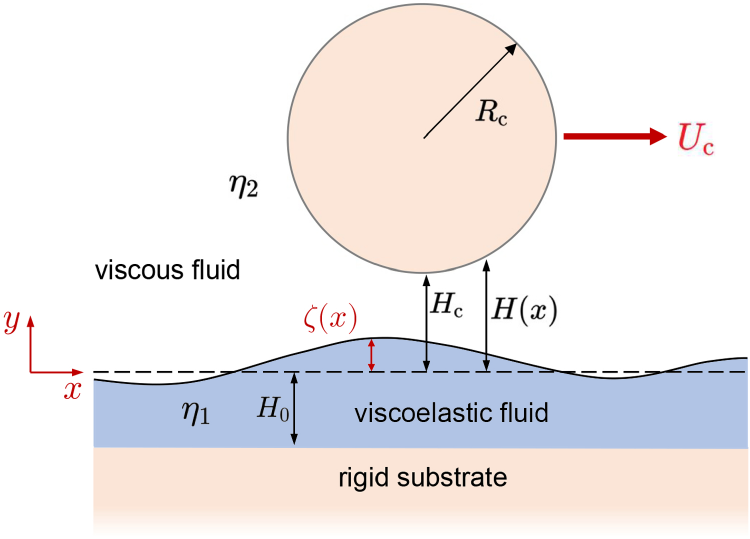}
\caption{A cylinder is translating parallel to a rigid substrate in a viscous fluid (called the outer fluid). The substrate is covered by a layer of viscoelastic Maxwell fluid of thickness $H_0$. The undeformed surface of the Maxwell fluid was located at $y=0$, but the motion of the cylinder deforms it to $y=\zeta(x,t)$. This deformation generates a lift force for the cylinder.}
\label{fig1}
\end{figure}

\subsection{Problem setting}\label{sec_2A}

We consider a cylinder of radius $R_{\mathrm{c}}$ moving with a constant velocity $U_{\mathrm{c}}$ above a viscoelastic fluid layer of thickness $H_0$ (see Fig.~\ref{fig1}). The cylinder is immersed in a viscous fluid (which we call the outer fluid) of viscosity $\eta_2$. We take a Cartesian coordinate $\{x,y\}$ along the horizontal and the vertical directions. The minimum separation between the cylinder and the undeformed interface is $H_{\mathrm{c}}$. At the limit $H_{\mathrm{c}} \ll R_{\mathrm{c}}$, the cylinder can be approximated by a parabola, and the surface profile of the cylinder is given by the equation
\begin{equation}
 H(x,t) = H_{\mathrm{c}}  \left[ 1+ \frac{(x-U_{\mathrm{c}}t)^2}{L^2}  \right] , \label{eq1}    
\end{equation}
where $L$ is the contact length defined by 
$L = \sqrt{2R_{\mathrm{c}}H_{\mathrm{c}}}$. In this paper, we consider the case of $H_{\mathrm{c}} \ll L$ and define the smallness parameter
\begin{equation}
\epsilon = \frac{H_{\mathrm{c}}}{L}
         = \sqrt{\frac{H_{\mathrm{c}}} 
              {2R_{\mathrm{c}}}}. \label{eq2}
\end{equation}

We assume that the layer is an incompressible viscoelastic fluid described by the Maxwell fluid model characterized by the viscosity $\eta_1$ and the relaxation time $\tau$. The complex shear modulus $G^*(\omega)$ is given by~\cite{Doi13}
\begin{equation}
G^*(\omega) = G_0\left[ \frac{(\omega\tau)^2}{1+(\omega\tau)^2} + i \frac{\omega\tau}{1+(\omega\tau)^2}\right], \label{eq3}
\end{equation}
where $G_0 = \eta_1/ \tau$. 

Let $\textbf{u}(\textbf{x},t)=(u_x, u_y)$ be the displacement vector of a material point located at position $\textbf{x} = (x,y)$ in the undeformed state. The normal displacement of the layer surface at $x$ is given by $\zeta(x) = u_y(x, y=0)$.

Each point in the layer fluid is displaced as the cylinder passes by. The duration time of the perturbation is given by $\tau_p = L/U_{\mathrm{c}}$.  We define a dimensionless parameter, the Deborah number, by
\begin{equation}
\mathrm{De} = \frac{\tau}{\tau_p} = \frac{\tau U_{\mathrm{c}}}{L}, \label{eq4}
\end{equation}
As $\mathrm{De} \to 0$, the viscoelastic fluid layer reduces to a viscous fluid layer with viscosity $\eta_1$. On the other hand, as $\mathrm{De} \to \infty$, it becomes a purely elastic solid layer with shear modulus $G_0$. 

\subsection{Scaling analysis of elastic layer}\label{sec_2B}

Before discussing the general case of viscoelastic fluid, we first discuss the case of purely elastic layer, i.e, the case of $\tau \to \infty$. Our purpose here is to show that the tangential shear stress, which has often been ignored in the previous analysis~\cite{Skotheim05, Kargar21}, is important for thin incompressible layer.

When the outer fluid is flowing, it exerts tangential stress $\sigma_{xy} \approx \eta_2 \partial v_{2}/\partial y$ and normal stress $\sigma_{yy}\approx p_2$ on the layer surface, where $v_{2}$ and $p_2$ are the tangential velocity and the pressure of the outer fluid. The magnitudes of the stresses can be estimated as~\cite{Skotheim04}
\begin{equation} \label{eq5}
\sigma_{xy} \sim \eta_2 \frac{U_{\mathrm{c}}}{H_{\mathrm{c}}},  \quad  
\sigma_{yy} \sim \eta_2 \frac{U_{\mathrm{c}} L}{H_{\mathrm{c}}^2}.
\end{equation}
Equation (\ref{eq5}) indicates that the tangential stress $\sigma_{xy}$ is much smaller than the normal stress $p_2$  by a factor of $\epsilon$. It is perhaps for this reason that the effect of tangential stress has been ignored in previous analysis. However, the smallness of the tangential stress does not mean that its effect is negligible for the deformation of the layer. 

An incompressible layer can only be deformed via shear with shear strain $u_x(x, 0)/H_0$. Equating the work done by the tangential stress and the normal stress with the elastic energy stored in the layer, we have
\begin{equation}\label{eq6}
\sigma_{xy} L u_x + p_2 L \zeta \sim G_0 \left(\frac{u_x}{H_0}\right)^2 H_0 L.
\end{equation}
On the other hand, for incompressible layer, we have 
$\nabla\cdot \textbf{u} = 0$, which gives  
$u_x/L \sim \zeta/H_0$. Together with Eqs.~(\ref{eq5}) and (\ref{eq6}), we obtain
\begin{gather}
\frac{u_x}{L} \sim \frac{\eta_2 U_{\mathrm{c}}}{G_0 L} \left(\frac{H_0}{H_{\mathrm{c}}} + \frac{H_0^2}{H_{\mathrm{c}}^2}\right) \label{eq7}, \\
\frac{\zeta}{H_{\mathrm{c}}} \sim \frac{\eta_2 U_{\mathrm{c}}}{G_0 L} \left(\frac{H_0^2}{H_{\mathrm{c}}^2} + \frac{H_0^3}{H_{\mathrm{c}}^3}\right). \label{eq8}
\end{gather}
The first terms in Eqs.~(\ref{eq7}) and (\ref{eq8}) represent the contributions of the tangential stress, and the second terms represent the contributions of the normal stress.

Equations~(\ref{eq7}) and (\ref{eq8}) indicate that the tangential force is as important as the normal force for $H_0 \sim H_{\mathrm{c}}$. This is because the tangential stress gives a larger contribution to the surface displacements than the normal stress by a factor of $H_{\mathrm{c}}/(H_0 \epsilon)$. As a result, the smallness parameter $\epsilon$ does not appear in Eqs.~(\ref{eq7}) and (\ref{eq8}). Therefore, for the normal stress to be the dominant effect, the layer should be thick with $H_0 \gg H_{\mathrm{c}}$. But for a thin incompressible layer of $H_{\mathrm{c}}/H_0 \sim 1 $, we need to account for both the normal stress and the tangential stress. In the following, we conduct such an analysis for an incompressible viscoelastic fluid layer. 

\subsection{Analysis for viscoelastic fluid layer}\label{sec_2C}

We now present our model that includes both the tangential and the normal stresses. The theory is valid ranging from thin ($H_{\mathrm{c}} \lesssim H_0 $) to thick ($H_{\mathrm{c}} \gg H_0$) viscoelastic layer. 

For small displacement, the stress-strain relation can be described by the linear viscoelastic model:
\begin{equation}\label{eq9}
\sigma_{jk}(\textbf{x}, t) = \int_{-\infty}^{t} G(t-t') 
\frac{\partial \epsilon_{jk}}{\partial t'}(\textbf{x}, t')\, dt' - p_1(\textbf{x}, t)\delta_{jk}.
\end{equation}
where $\sigma_{jk}(t)$ is the $j$-$k$ component of the stress tensor, $\epsilon_{jk}(x,t)=\partial u_j/\partial x_k + \partial u_k/\partial x_j$ is the the $j$-$k$ component of the strain tensor, $G(t)$ is the relaxation modulus, and $p_1$ is the pressure in the viscoelastic layer.

Using the time Fourier transform, 
\begin{equation}\label{eq10}
f(\omega) = \int_{-\infty}^{\infty}f(t)e^{-i\omega t}\,dt,
\end{equation}
we can rewrite Eq.~(\ref{eq9}) as
\begin{equation}\label{eq11}
\sigma_{jk}(\textbf{x}, \omega) = G^{\ast}(\omega) \epsilon_{jk}(\textbf{x}, \omega) - p_1(\textbf{x}, \omega)\delta_{jk},
\end{equation}
where $G^{\ast}(\omega) = i\omega \int_{0}^{\infty} G(t) e^{-i\omega t}\,dt$. 

A differential equation for the normal displacement $u_y$ can be derived by introducing the space Fourier transform~\cite{Skotheim05},
\begin{equation}\label{eq12}
\tilde{f}(k) = \int_{-\infty}^{\infty} f(x) e^{-ikx}\,dx.
\end{equation}
Since the cylinder is translating with a constant velocity $U_{\mathrm{c}}$, the steady-state quantities must have the form of a travelling wave, $f(x,t) = f_0(x-Ut)$, and its Fourier transform is written as
\begin{equation}\label{eq13}
\tilde{f}(k,\omega) = 2\pi \delta(\omega + Uk) \tilde{f}_0(k)
\end{equation}
Taking the transform from $\omega$ to $t$ yields $\tilde{f}(k, t) = e^{-i k U_{\mathrm{c}} t}\tilde{f}_0(k)$, where the phase factor $e^{-ikUt}$ captures the only time-dependent part and describes the translation with the cylinder. Therefore, the stress-strain relation in steady state can be written as~\cite{Long96, Karpitschka15, Pandey16}
\begin{equation}\label{eq14}
\tilde{\sigma}_{jk}(k, y) = G^{\ast}(-Uk) \tilde{\epsilon}_{jk}(k, y) - \tilde{p}_1(k, y)\delta_{jk},
\end{equation}

The force balance equation in the viscoelastic layer, $\nabla\cdot \boldsymbol{\sigma} = 0$, is written as
\begin{gather}
-i k \tilde{p}_1 + G^{\ast}\left(-k^2 \tilde{u}_x + \frac{\partial^2 \tilde{u}_x}{\partial y^2}\right) = 0, \label{eq15} \\
-\frac{\partial \tilde{p}_1}{\partial y} + G^{\ast}\left(-k^2 \tilde{u}_y +\frac{\partial^2 \tilde{u}_y}{\partial y^2} \right) = 0, \label{eq16}
\end{gather}
The incompressible condition is written as
\begin{equation}\label{eq17}
ik\tilde{u}_x 
+ \frac{\partial \tilde{u}_y}{\partial y}  = 0.
\end{equation}
From Eqs. (\ref{eq15})--(\ref{eq17}), it follows that
\begin{equation}\label{eq18}
\frac{\partial^4 \tilde{u}_y}{\partial y^4} - 2k^2 \frac{\partial^2 \tilde{u}_y}{\partial y^2} + k^4 \tilde{u}_y = 0,
\end{equation}
and
\begin{equation}\label{eq19}
\tilde{p}_1 = -\frac{G^{\ast}}{k^2}\left(k^2\frac{\partial \tilde{u}_y}{\partial y} - \frac{\partial^3 \tilde{u}_y}{\partial y^3}\right).
\end{equation}
Equation~(\ref{eq18}) can be solved for $\tilde{u}_y(k,y)$ if the boundary conditions at $y=-H_0$ and $y=0$ are given.

The boundary conditions at the substrate surface, $y = -H_0$, are given by $\tilde{u}_x=\tilde{u}_y=0$. By 
Eq.~(\ref{eq17}), they give the following condition for
$\tilde{u}_y$
\begin{equation}\label{eq20}
\tilde{u}_y =0, \quad  \frac{\partial \tilde{u}_y}{\partial y} = 0 \quad  \text{at } y=-H_0. 
\end{equation}

\footnotetext[1]{Writing down this requires careful consideration since the coordinate $(x,y)$ in the viscoelastic layer stands for the position of the point in the undeformed layer. The point $(x,0)$ on the interface is displaced to $(x+u_x, \zeta)$. The tangential displacement $u_x \sim L/H_0 \zeta$ by the imcompressibility condition. For $\zeta \ll H_0$, $u_x \ll L$. Therefore, we assume that the interface is located at $(x, 0)$ in the boundary conditions.}

The boundary condition at the interface between the viscoelastic layer and the outer fluid, $y=0$, is given by the force balance conditions~\cite{Note1}.  Since 
the tangential stress is continuous at the interface, we have
\begin{gather}\label{eq21} 
G^{\ast}\left(-\frac{1}{ik}\frac{\partial^2 \tilde{u}_y}{\partial y^2} + ik\tilde{u}_y\right) 
= \eta_2 \frac{\partial \tilde{v}_2}{\partial y} 
\end{gather}
The normal stress has a jump $\gamma \partial^2 \zeta/\partial x^2$ at the interface due to the surface tension $\gamma$. Therefore, the force-balance condition for normal direction is written as
\begin{equation}\label{eq22}
-\tilde{p}_1 + 2G^{\ast}\frac{\partial \tilde{u}_y}{\partial y} = -\tilde{p}_2 - \gamma k^2\tilde{\zeta}.
\end{equation}
Substituting Eq.~(\ref{eq19}), Eq.~(\ref{eq22}) becomes
\begin{equation}\label{eq23}
3G^{\ast}\frac{\partial \tilde{u}_y}{\partial y} - \frac{G^{\ast}}{k^2}\frac{\partial^3 \tilde{u}_y}{\partial y^3} = -\tilde{p}_2 - \gamma k^2\tilde{\zeta}.
\end{equation}

To complete the boundary condition at $y=0$, we need to know $\tilde{v}_2$ in Eq.~(\ref{eq21}). This is given by solving the Stokes equations of the outer fluid, 
\begin{equation}\label{eq24}
\eta_2 \frac{\partial^2 v_2}{\partial y^2} = \frac{\partial p_2}{\partial x},
\end{equation}
The boundary conditions for $v_2$ are $v_2(x,y) = U_{\mathrm{c}}$ at $y = H(x)$ and $v_2(x,y) = U_{\mathrm{s}}(x)$ at $y=0$, where $U_{\mathrm{s}}(x)$ is the tangential velocity (or the slip velocity) at the layer surface. By definition, $U_{\mathrm{s}} = \partial_t u_x$ at $y = 0$. Since $u_x$ has the form of $u_x(x-U_c t, y)$, we have
\begin{equation}\label{eq25}
U_{\mathrm{s}}(x) = -U_{\mathrm{c}}\frac{\partial u_x}{\partial x}(x, 0) = U_{\mathrm{c}}\frac{\partial u_y}{\partial y}(x, 0). 
\end{equation}
The solution of Eq.~(\ref{eq24}) is given by
\begin{equation}\label{eq26}
v_2(x,y) = \frac{1}{2\eta_2} \frac{\partial p_2}{\partial x}y[y-H(x)] + \frac{U_{\mathrm{c}}-U_{\mathrm{s}}(x)}{H(x)}y + U_{\mathrm{s}}(x). 
\end{equation}
Substituting into Eq.~(\ref{eq21}), we obtain
\begin{equation}\label{eq27}
\begin{aligned}
G^{\ast} \left(-\frac{1}{ik}\frac{\partial^2 \tilde{u}_y}{\partial y^2} + ik\tilde{u}_y \right) = \eta_2 \big(U_{\mathrm{c}}\widetilde{H^{-1}} - \\
\tilde{U}_{\mathrm{s}} \ast \widetilde{H^{-1}}\big) - \frac{1}{2} (ik\tilde{p}_2) \ast \tilde{H},
\end{aligned}
\end{equation}
where the convolution in Fourier space is defined as 
\begin{equation}\label{eq28}
\tilde{f}_1 \ast \tilde{f}_2 = \frac{1}{2\pi}\int_{-\infty}^{\infty} \tilde{f}_1 (k') \tilde{f}_2 (k-k')\,dk'.
\end{equation}

The remaining unknown function is $p_2(x)$. The pressure $p_2(x)$ is determined by the condition that the flux in the channel is constant and satisfies the following Reynolds equation~\cite{Hamrock04},
\begin{equation}\label{eq29}
\frac{\partial}{\partial x}\left[
\frac{1}{\eta_2} h_2^3\frac{\partial p_2}{\partial x} + 6 h_2(U_{\textrm{c}}-U_{\mathrm{s}})
\right] = 0,
\end{equation}
where the gap thickness between the cylinder and the deformed interface $h_2(x) = H(x)-\zeta(x)$. Notice that in the calculation of $p_2(x)$, the interface slip velocity $U_{\mathrm{s}}(x)$ is included in Eq.~(\ref{eq29}). The slip velocity given by Eq.~(\ref{eq25}) is a direct consequence of the tangential displacement of the interface and must be included.  

Equations~(\ref{eq18}), (\ref{eq20}), (\ref{eq23}), (\ref{eq25}), (\ref{eq27}), and (\ref{eq29}) are complete for determining $u_y$, $U_{\mathrm{s}}$, and $p_2$. Given $p_2$, the lift force per unit length upon the cylinder is computed by 
\begin{equation}\label{eq30}
F = \int_{-\infty}^{\infty} p_2\,dx.
\end{equation}

\subsection{Numerical methods}\label{sec_2D}
We solved the above set of equations numerically. To make the equations non-dimensional, we scale $x$ with the contact length $L$, $y$ and $H_0$ with $H_{\mathrm{c}}$, time with the contact time $\tau_p$ and pressure with the lubrication pressure 
\begin{equation}\label{eq31}
p_0 = \frac{\eta_2 U_{\mathrm{c}} L}{H_{\mathrm{c}}^2} .
\end{equation} 
The two dimensionless control parameters are the Deborah number De [Eq.~(\ref{eq4})] and the capillary number, defined by
\begin{equation}\label{eq32}
\mathrm{Ca} = \frac{\eta_2 U_{\mathrm{c}}}{\gamma }
\end{equation}

To solve the coupled equations for the layer fluid and the outer fluid, we adopt a numerical iteration scheme. In each iteration, a guess solution to $p_2(x)$ is used to calculate the deformation $\zeta(x)$ using the MATLAB routine \verb!bvp4c! and the slip velocity $U_{\mathrm{s}}(x)$, which are then used in Eq.~(\ref{eq27}) to update $p_2(x)$. The iteration is stopped when the update for $\zeta(x)$ is smaller than $10^{-7}$. Throughout our numerical calculations, we set the value of the smallness parameter $\epsilon = 0.1$. 

As the initial guess for $p_2(x)$, we take the pressure $p_2^{(0)}$ for the undeformed interface, which satisfies
\begin{equation}\label{eq33}
\frac{\partial}{\partial x} \left[\frac{1}{\eta_2}H^3\frac{\partial p_2^{(0)}}{\partial x} + 6 H U_{\textrm{c}}\right] = 0. 
\end{equation}
$p_2^{(0)}$ is given by
\begin{equation}\label{eq34}
 p_2^{(0)}(x) = \frac{2 p_0 L^3 x}{(L^2 + x^2)^2}. 
\end{equation} 

\section{Results}\label{sec_3}
\begin{figure*}[t]
\centering
\includegraphics[bb=0 5 360 75, scale=1.4, draft=false]{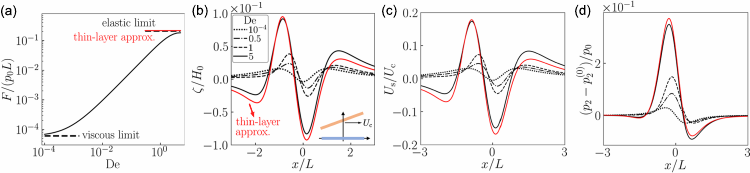}
\caption{Effect of Deborah number De with $\mu = 0.1$, $\mathrm{Ca} = 0.1$, and $H_0/H_{\mathrm{c}} = 0.3$. The red lines are the numerical results of the thin-layer approximation. (a) Lift force versus De for $10^{-4} < \text{De} < 5$. The dark dashed lines indicate the viscous and elastic limits of our full model. (b) Deformation $\zeta$, (c) slip velocity $U_{\mathrm{s}}$, and (d) disturbed pressure $p_2-p_2^{(0)}$ for increasing $\mathrm{De}$ from $10^{-4}$ to 5. Inset in (b) shows the simplified geometry near the close-contact region for $\mathrm{De} = 5$.}
\label{fig2}
\end{figure*}

\subsection{Numerical Solution}\label{sec_3A}

Figure~\ref{fig2} shows the results of the numerical calculation for the full model. Here, the effect of the Deborah number De is shown and all the other parameters are fixed as follows: the layer thickness ratio $H_0/H_{\mathrm{c}} = 0.3$, the capillary number $\mathrm{Ca}=0.1$ (surface tension is negligible), and the viscosity ratio $\mu = \eta_2/\eta_1 = 0.1$. 

The lift force $F$ as a function of the Deborah number De is shown in Fig.~\ref{fig2}(a). As De increases, the lift force increases continuously from the viscous limit to the elastic limit. The elastic limit is calculated by setting $G^{\ast}=G_0$, and the viscous limit is calculated by setting $G^{\ast} = i\omega\eta_1$. 

Figures~\ref{fig2}(b) and \ref{fig2}(c) show the interface profile $\zeta(x)$ and the slip velocity $U_{\mathrm{s}}(x)$, which appear in the the Reynolds equation [Eq.~\ref{eq29}]. As De increase, both $\zeta(x)$ and $U_{\mathrm{s}}(x)$ change from a nearly symmetric shape to an asymmetric shape.  

Figure~\ref{fig2}(d) shows the disturbed pressure $p_2(x)-p_2^{(0)}(x)$. Here, since $p_2(x)$ is close to $p_2^{(0)}(x)$ [Eq.~(\ref{eq34})], their difference is shown. Since the integral of $p_2^{(0)}(x)$ from $-\infty$ to $\infty$ is zero, the integral of the curve in Fig.~\ref{fig2}(d) gives the lift force. As it is seen in Fig.~\ref{fig2}(d), the disturbed pressure $p_2-p_2^{(0)}$ morphs from a nearly antisymmetric shape to a strongly asymmetric shape.  This gives the increase of the lift force shown in Fig.~\ref{fig2}(a).

\subsection{Elastic layer limit}\label{sec_3B}

\subsubsection{Perturbation calculation of the thin-layer approximation}\label{sec_3B1}

We now discuss the elastic limit in more detail. First, we consider the deformation of a thin incompressible elastic layer 
subjected to the tangential stress $\sigma_{xy}(x)$ and the
normal stress $p_2(x)$ applied on the interface by the outer fluid.
At the lubrication limit, $H_0 \ll L$, the tangential displacement $u_x(x,y)$ satisfies the following equation, 
\begin{equation}\label{eq35}
G_0 \frac{\partial^2 u_x}{\partial y^2} = \frac{\partial p_1}{\partial x}.
\end{equation}
Here, $p_1 = p_2$ due to the continuity in the normal stress at the interface. The solution to Eq.~(\ref{eq35}) is 
\begin{equation}\label{eq36}
u_x (x,y) = \frac{1}{2G_0}\frac{\partial p_2}{\partial x} (y^2-H_0^2) + \frac{y+H_0}{G_0}\sigma_{xy}(x).
\end{equation}
The normal displacement $u_y(x,y)$ can be obtained by integrating the incompressible condition $\partial u_x/\partial x +\partial u_y/\partial y=0$. Therefore, the displacements at the surface are given by 
\begin{gather}
u_x (x, 0) = \frac{H_0}{G_0}\sigma_{xy}(x) - \frac{H_0^2}{2G_0}\frac{\partial p_2}{\partial x}(x), \label{eq37}  \\
\zeta(x) = -\frac{H_0^2}{2G_0} \frac{\partial\sigma_{xy}}{\partial x}(x) + \frac{H_0^3}{3 G_0}\frac{\partial^2 p_2}{\partial x^2}(x). \label{eq38}
\end{gather}
Equations~(\ref{eq37}) and (\ref{eq38}) have been derived previously by~\citet{Chandler20} through an asymptotic expansion in $H_0/L$. Equations~(\ref{eq37}) and (\ref{eq38}) together with the Reynolds equation [Eq.~(\ref{eq29})] and Eq.~(\ref{eq25}) allow us to calculate $p_2(x)$ and the lift force. These equations can be regarded as an approximation for the elastic limit of the full model for $H_0 \ll L$. Therefore, we shall call this approach the thin-layer approximation. 

We solved the set of equations given by the
thin-layer approximation by perturbation method.
We define a dimensionless parameter $\alpha$ by
\begin{equation}\label{eq39}
\alpha = \frac{\eta_2 U_{\textrm{c}}}{G_0 L}. 
\end{equation}
and solve the equations regarding $\alpha$ as a small parameter. As shown by Eqs.~(\ref{eq7}) and (\ref{eq8}), $\zeta/H_{\mathrm{c}} \sim \alpha$ and $U_{\textrm{s}}/U_{\textrm{c}} \sim \alpha$. Therefore, we can expand $p_2$ as $p_2 = p_2^{(0)}+p_2^{(1)}$, where $p_2^{(0)}$ is the solution for the undeformed interface [Eq.~(\ref{eq34})]. The disturbed pressure $p_2^{(1)}$ due to $\zeta$ and $U_{\textrm{s}}$ satisfies the first-order equation,   
\begin{equation}\label{eq40}
\begin{aligned}
\frac{\partial}{\partial x}\left[\frac{1}{\eta_2}\left(H^3\frac{\partial p_2^{(1)}}{\partial x} - 3H^2\zeta\frac{\partial p_2^{(0)}}{\partial x}\right) - 6U_{\textrm{c}}\zeta - 6U_{\textrm{s}}H\right] = 0.
\end{aligned}
\end{equation}
We first calculate $\zeta$ and $U_{\textrm{s}}$ by substituting $p_2^{(0)}$ into Eqs.~(\ref{eq37}), (\ref{eq38}), and (\ref{eq25}) and then solve Eq.~(\ref{eq40}) for $p_2^{(1)}$. The lift force obtained by integrating $p_2^{(1)}$ is
\begin{gather}\label{eq41}
F = \pi \frac{\eta_2^2 U_{\mathrm{c}}^2 L}{G_0 H_{\mathrm{c}}^2} \left(\frac{3}{4}\frac{H_0}{H_{\mathrm{c}}} + \frac{27}{16}\frac{H_0^2}{H_{\mathrm{c}}^2} + \frac{15}{16}\frac{H_0^3}{H_{\mathrm{c}}^3}\right).
\end{gather}
Here, the $H_0^3$ scaling represents the contribution by the deformation caused by $p_2$ and has been discussed in Refs.~\cite{Skotheim04, Skotheim05, Kargar21}. Equation (\ref{eq41}) indicates that there are two other terms at the leading orders in $H_0$: the $H_0$ term is the contribution of the slip velocity caused by the tangential stress; the $H_0^2$ term is the combined effect of the deformation due to the tangential stress and the slip velocity due to the pressure.

\subsubsection{Dependence on $\alpha$}\label{sec_3B2}
\begin{figure}[t]
\centering
\includegraphics[bb=0 10 362 292, scale=0.66,draft=false]{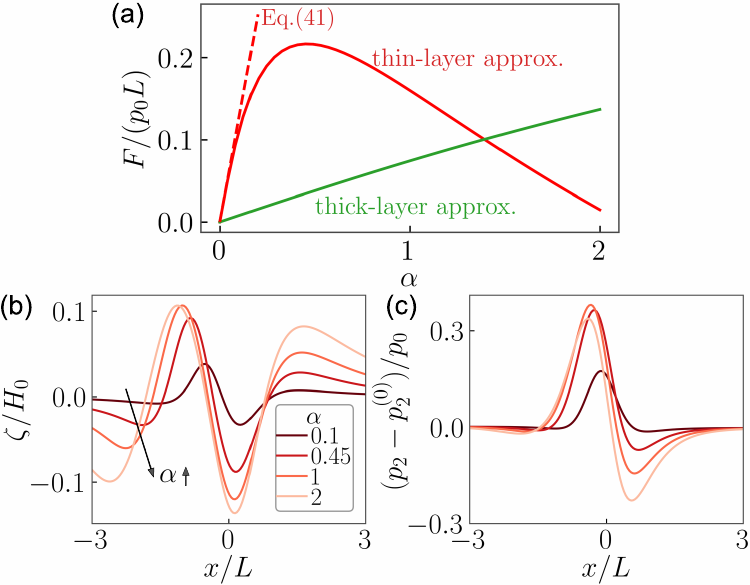}
\caption{Effect of $\alpha$ for a thin elastic layer of $H_0/H_{\mathrm{c}} = 0.3$. (a) Lift force as a function of $\alpha$ for the thin-layer approximation (red curve) and thick-layer approximation (green curve). Red dashed line is the asymptotic solution for small $\alpha$ given by Eq.~(\ref{eq41}). (b) $\zeta$ and (c) $p_2-p_2^{(0)}$ for different values of $\alpha$.}
\label{fig3}
\end{figure}

When $\alpha$ is not small, the set of equations of thin-layer approximation has to be solved numerically. We adopt the same iteration scheme as in Sec.~\ref{sec_2D}. In each iteration, a guess solution to $p_2$ is substituted into Eqs.~(\ref{eq37}), (\ref{eq38}), and (\ref{eq25}) to obtain $\zeta$ and $U_{\mathrm{s}}$. The solutions are used to update $p_2$ by solving Eq.~(\ref{eq29}). The results of such calculation
for $\alpha = 0.5$ are shown in Fig.~\ref{fig2} with red solid lines. As shown in Fig.~\ref{fig2}(a), the lift force obtained using the thin-layer approximation overlaps with the full model at the elastic limit. Figures~\ref{fig2}(b), (c), and (d) show that the surface displacement $\zeta$, the slip velocity $U_{\mathrm{s}}$, and the disturbed pressure $p_2-p_2^{(0)}$ all agree well with the full model for $\mathrm{De} = 5$. Therefore, for $H_0 \lesssim H_{\mathrm{c}}$, our full model is well approximated by the thin-layer approximation.

The dependence of $F$ on $\alpha$ is shown in Fig.~\ref{fig3}(a). As $\alpha$ increases, the magnitude of $\zeta$ increases but the profile becomes less antisymmetric with an arising local maximum [Fig.~\ref{fig3}(b)]. As a result, the magnitude of the disturbed pressure $p_2-p_2^{(0)}$ increases but it becomes more antisymmetric as $\alpha$ increases [Fig.~\ref{fig3}(c)]. These two competing effects produce a maximum lift force when $\alpha \approx 0.45$, which indicates an optimal choice of the shear modulus $G_0$. The green solid line shows the result obtained from the thick-layer approximation that will be discussed in the next section.

\subsubsection{Dependence on $H_0$}\label{sec_3B3}

\begin{figure}[t]
\centering
\includegraphics[bb=0 10 365 235, scale=0.66,draft=false]{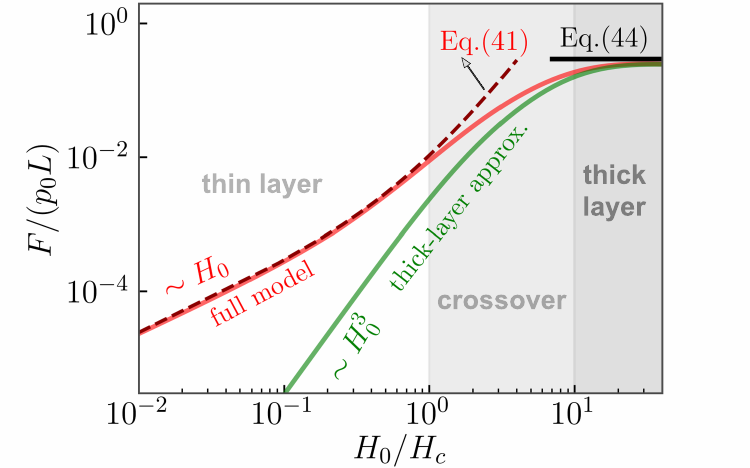}
\caption{Lift force $F$ versus layer thickness $H_0$ with $\alpha = 10^{-3}$. Red curve is the numerical result of the full model at the elastic limit. Red dashed line is the asymptotic solution given by Eq.~(\ref{eq41}). Green curve is the numerical result of the thick-layer approximation. Dark solid line is the asymptotic solution at the limit $H_0 \to \infty$ given by Eq.~(\ref{eq44}).}
\label{fig4}
\end{figure}

We showed in section \ref{sec_3B1} that the tangential stress and the slip velocity affect the lift force greatly for $H_0 \lesssim H_{\mathrm{c}}$. To explore the dependence for a wide range of $H_0$, we numerically calculated the lift force using the full model at the elastic limit ($G^{\ast} = G_0$). The result for $\alpha=10^{-3}$ is shown in Fig.~\ref{fig4}. Here, the numerical result is denoted by the red solid line, and the asymptotic result given by Eq.~(\ref{eq41}) is denoted by the red dashed line. For $H_0 \lesssim H_{\mathrm{c}}$, the asymptotic solution is a good approximation to the full model. The lift force $F$ scales as $H_0$ for $H_0 \ll H_{\mathrm{c}}$. For a thick layer of $H_0 \gg H_{\mathrm{c}}$, $F$ approaches a constant value. 

\footnotetext[2]{In Ref.~\cite{Kargar21}, the deformation of an elastic layer due to the normal stress $p_2$ is computed using the thick-layer approximation, but the lift force is calculated based on the Lorentz reciprocal theorem.}

Previous works on incompressible elastic layer ignore the effect of tangential stress for both thin and thick layers and only consider the deformation due to the normal stress.  In such a treatment, the deformation is generally expressed by the Green's function $\mathcal{G}$ for the point normal force~\cite{Skotheim05, Kargar21},
\begin{equation}\label{eq42}
\zeta(x) = \int \mathcal{G}(x-x')p_2(x')\,dx'.
\end{equation}
The Green's function $\mathcal{G}$ in Fourier space can be written as
\begin{equation}\label{eq43}
\tilde{\mathcal{G}}(k) = -\frac{1}{G_0} \frac{-2kH_0 + \sinh(2kH_0)}{2k [1+\cosh(2k H_0)+ 2k^2 H_0^2]}.
\end{equation}

To examine the validity of the previous treatment, we calculated the deformation and lift force by solving Eqs.~(\ref{eq29}), (\ref{eq42}), and (\ref{eq43}) (with $U_{\mathrm{s}} = 0$) using the same numerical iteration scheme as the full model~\cite{Note2}. The dependence of $F$ on $H_0$ is shown in Fig.~\ref{fig4} by the green solid line. It is seen that for $H_0 \gg H_{\mathrm{c}}$, the previous treatment gives results close to that of the full model, but for $H_0 \lesssim H_{\mathrm{c}}$, significantly difference is seen. Therefore, the previous treatment is only valid for a thick layer. Hence we call the previous analysis the thick-layer approximation.

Figure~\ref{fig4} indicates that for large $H_0$ the lift force approaches to a constant value independent of $H_0$. This can be shown analytically. In this limit, the Green's function is given by $\tilde{\mathcal{G}}\approx -1/(2 G_0 |k|)$. Following the same perturbation method as the thin-layer approximation, we first compute $\zeta$ using $p_2^{(0)}$ and then compute $p_2^{(1)}$ using Eq.~(\ref{eq40}). The resulting lift force is~\cite{Skotheim05, Kargar21}
\begin{equation}\label{eq44}
F_{H_0 \to \infty} = \frac{3\pi}{32}\frac{\eta_2^2 U_{\mathrm{c}}^2 L^4}{G_0 H_{\mathrm{c}}^5} = \frac{3\pi}{8}\frac{\eta_2^2 U_{\mathrm{c}}^2 R_{\mathrm{c}}^2}{G_0 H_{\mathrm{c}}^3}.
\end{equation}
The value calculated by Eq.~(\ref{eq44}) is shown in Fig.~\ref{fig4} with the dark solid line. Both the full model and the thick-layer approximation approach to the same value of Eq.~(\ref{eq44}) at large $H_0$. 

Although our full model and the previous treatment (thick-layer approximation) agree well in thick layer, their deviation becomes significant in thin layer. Figure~\ref{fig3}(a) shows the lift force calculated by our full model with the red line, and that by the previous treatment with the green line. The previous treatment predicts that the lift force increases monotonically with $\alpha$, while our full model predicts that the lift force has a maximum, i.e., there is an optimal elastic modulus to give the largest lift force. The existence of the optimal modulus has been reported by the thick-layer approximation in thick layer~\cite{Skotheim05}. Our calculation indicates that such optimal modulus also exists in thin layer. 

Therefore, we conclude that the tangential stress gives important contributions to the lift force for a thin layer of $H_0 \lesssim H_{\mathrm{c}}$ and is even dominant for $H_0 \ll H_{\mathrm{c}}$. For a thick layer of $H_0 \gg H_{\mathrm{c}}$, the normal stress $p_2$ dominates and the tangential stress can be ignored. In the intermediate regime, both the tangential and normal stresses should be taken into account. 

Finally, we note that the discussion here is limited to incompressible layer. If the layer is compressible, the previous treatment that only considers the deformation caused by the normal stress will be valid even for thin layer of $H_0 \lesssim H_{\mathrm{c}}$.

\subsection{Viscous fluid limit}\label{sec_3C}

\subsubsection{Numerical solution}\label{sec_3C1}
\begin{figure}[b]
\centering
\includegraphics[bb=0 12 365 132, scale=0.66,draft=false]{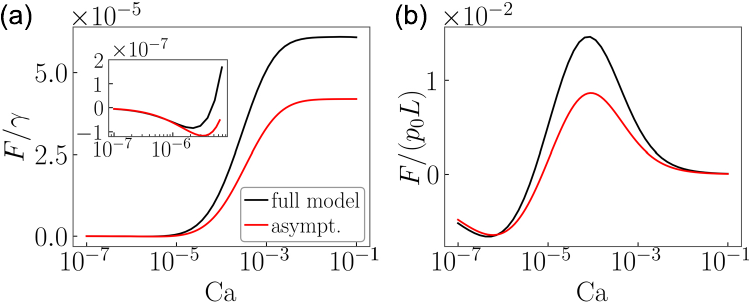}
\caption{Lift force for a thin viscous fluid layer with $H_0/H_{\mathrm{c}} = 0.3$ and $\mu = 0.1$. The dark curve is the numerical solution of the full model with $G^{\ast} = i\omega\eta_1$. The red curve is the asymptotic solution based on Stokesian hydrodynamics (see text in Sec.~\ref{sec_3C3}). (a) $F/\gamma$ versus Ca. The inset shows a zoomed-in view for the region $\text{Ca} \lesssim 10^{-5}$. (b) $F/p_0 L$ versus Ca.}
\label{fig5}
\end{figure}

We now consider the other limit, the fluid layer limit where $\mathrm{De} \to 0$.  In this case, the control parameter is the capillary number Ca, which can be changed by varying either the translating velocity $U_{\mathrm{c}}$ or the surface tension $\gamma$.

Numerical solution for this limit is obtained by setting $G^{\ast} = i\omega \eta_1$. Figure~\ref{fig5} shows the result of such calculation. For $\text{Ca} \lesssim 10^{-5}$, the lift force is close to zero. The lift force increases with Ca for $10^{-5} \lesssim \text{Ca} \lesssim 10^{-2}$, and saturates for $\text{Ca} \gtrsim 10^{-2}$. 

For small Ca, a zoomed-in view shows that although small, $F/\gamma$ is actually negative for $\text{Ca} \lesssim 10^{-5}$ [Fig.~\ref{fig5}(a) inset]. The negative lift force at low Ca becomes more noticeable when $F/(p_0 L)$ is plotted against Ca [see Fig.~\ref{fig5}(b)]. 

\begin{figure}[t]
\centering
\includegraphics[bb=0 12 365 290, scale=0.66,draft=false]{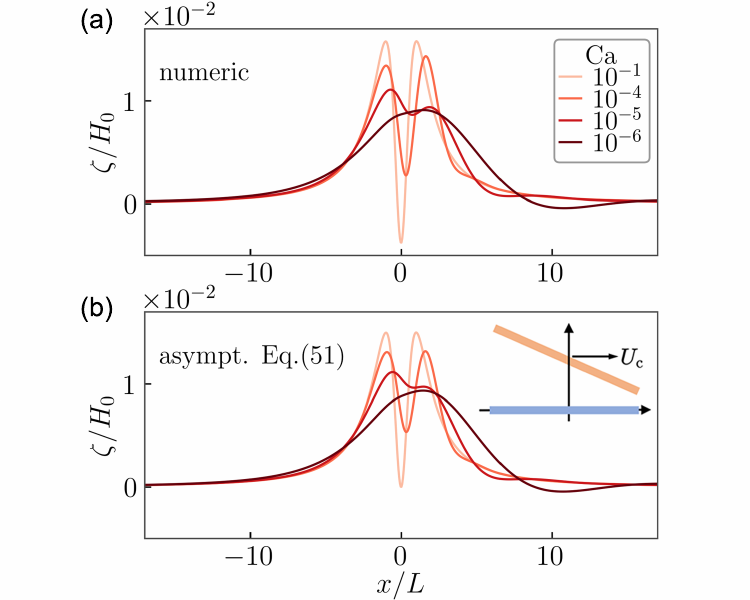}
\caption{Effect of Ca on $\zeta$ obtained from (a) numerical calculation of the full model and (b) asymptotic calculation. Inset in (b) shows the simplified geometry near the closest-contact region for $\textrm{Ca} = 10^{-6}$.}
\label{fig6}
\end{figure}

Figure~\ref{fig6}(a) shows the interface profile $\zeta(x)$ for the fluid layer, which looks very different from the elastic case. For large Ca, $\zeta(x)$ is nearly symmetric with two peaks. As Ca is decreased, the two peaks come close to each other and eventually merge into one at $\text{Ca} = 10^{-6}$. 

\subsubsection{Scaling analysis}\label{sec_3C2}

To gain insights, we first estimate the magnitude of the surface deformation $\zeta$ using scaling analysis. The work done by the tangential and the normal stresses of the outer fluid layer is balanced with the dissipated energy and the interface energy,
\begin{equation}\label{eq45}
\eta_2\frac{U_{\mathrm{c}}}{H_{\mathrm{c}}} L U_{\mathrm{s}}\tau_p + p_0 L \zeta \sim \eta_1 \left(\frac{U_{\mathrm{s}}}{H_0}\right)^2 H_0 L \tau_p + \gamma \frac{\zeta^2}{L}.
\end{equation}
To calculate $\zeta$, we need to write $U_{\mathrm{s}}$ in terms of $\zeta$. From Eq.~(\ref{eq25}), $U_{\mathrm{s}} \sim U_{\mathrm{c}}\zeta/H_0$. Substituting into Eq.~(\ref{eq45}), we obtain
\begin{equation}\label{eq46}
\frac{\zeta}{H_{\mathrm{c}}} = \mu \left(\frac{H_0^2}{H_{\mathrm{c}}^2} + \frac{H_0^3}{H_{\mathrm{c}}^3}\right)\left(1+\text{Ca}^{-1}\mu H_0^3/L^3\right)^{-1}.
\end{equation}
Here, the $H_0^2$ term is the contribution of the tangential stress and the $H_0^3$ term is the contribution of the normal stress. This scaling relation of $\zeta$ shows a similar dependent on $H_0$ to the elastic case [Eq.~(\ref{eq8})], suggesting that the tangential stress is also important for a thin viscous layer.

\subsubsection{Perturbation calculation based on Stokesian hydrodynamics}\label{sec_3C3}

To validate the numerical solution of the full model at the limit $\mathrm{De}\to 0$, we derive the differential equation satisfied by $\zeta$ using the variational principle in Stokesian hydrodynamics~\cite{Happel12, Doi15}. Using lubrication approximation in both layers, we first write the energy dissipation function (per unit area),
\begin{equation}\label{eq47}
\Phi(x) = \frac{\eta_1}{2}\int \left(\frac{\partial v_1}{\partial y}\right)^2 \,dy + \frac{\eta_2}{2}\int \left(\frac{\partial v_2}{\partial y}\right)^2\, dy.
\end{equation}
We can further express $\Phi$ using the height-averaged velocity $\bar{v}_1$ and $\bar{v}_2$ (see Appendix~\ref{app_A}). The free energy (per unit area) due to the surface tension is 
\begin{equation}\label{eq48}
A = \gamma \left[1 + \frac{1}{2}\left(\frac{\partial\zeta}{\partial x}\right)^2 \right]. 
\end{equation}
The Rayleighian is then constructed as 
\begin{equation}\label{eq49}
\mathcal{R} = \int_{-\infty}^{\infty} \left[\Phi+ \dot{A} - p \frac{\partial}{\partial x}(\bar{v}_1 h_1 + \bar{v}_2 h_2)\right]\,dx, 
\end{equation}
where the average pressure $p$ is used to enforce the constraint of volume conservation. We solve $\bar{v}_1$ and $\bar{v}_2$ using the minimization conditions, $\delta \mathcal{R}/\delta \bar{v}_1 = 0$ and $\delta \mathcal{R}/\delta \bar{v}_2 = 0$. The coupled equations of $\zeta$ and $p$ are given by
\begin{equation}\label{eq50}
\frac{\partial}{\partial x}(\bar{v}_1 h_1) = 0,\quad \frac{\partial}{\partial x}(\bar{v}_1 h_1 + \bar{v}_2 h_2) = 0.
\end{equation}

For small deformation, we solve the obtained equations by perturbation. We expand $p$ as $p \approx p_2^{(0)} + p^{(1)}$. The perturbation parameter $\beta$ is given by the characteristic deformation size arising from Eq.~(\ref{eq46}), $\beta = \mu H_0^2/H_{\mathrm{c}}^2$. The zeroth-order equation in $\beta$ is simply Eq.~(\ref{eq33}) and the first-order equations are given in Appendix~\ref{app_A}. Keeping the leading order terms of $H_0$ and substituting $p_2^{(0)}$, we obtain
\begin{equation}\label{eq51}
\frac{H_0^3}{3 \text{Ca}}\frac{\partial^4\zeta}{\partial x^4} - \frac{\partial\zeta}{\partial x} = 4\mu \frac{H_0^2 L^2}{H_{\mathrm{c}}}\frac{x(x^2-L^2)}{(x^2+L^2)^3}.
\end{equation}
The first term here resembles the restoring bending force of an elastic plate of thickness $H_0$. A larger Ca corresponds to a smaller Young's modulus~\cite{Landau86}. We identify that the term on the right hand side represents the effect of viscous shear stress. For $\text{Ca} \gg \epsilon^3 (H_0/H_{\mathrm{c}})^3$, the first term is ignored and the second term $\partial \zeta/\partial x$ gives a symmetric deformation directly caused by the shear stress. However, the pressure-induced deformation is not included in Eq.~(\ref{eq51}) and is captured by terms at the next order, as indicated by Eq.~(\ref{eq46}). As shown in Fig.~\ref{fig6}(b), the asymptotic solution agrees well with the numerical result [Fig.~\ref{fig6}(a)], especially for small $\text{Ca} = 10^{-6}$. The lift force, calculated by integrating $p^{(1)}$, follows the same trend as the numerical solution (Fig.~\ref{fig5}). The value difference may be attributed to the contribution from $p_2$ that is not captured by the asymptotic solution. 

\begin{figure}[t]
\centering
\includegraphics[bb=0 12 365 130, scale=0.66,draft=false]{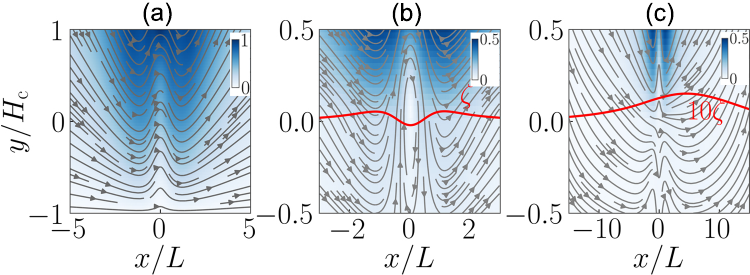}
\caption{The flow field in steady state with $H_0 = H_{\mathrm{c}}$ from numerical calculations. The solid substrate is located at $y/H_{\mathrm{c}} = -1$ and the bottom of the cylinder is located at $y/H_{\mathrm{c}} = 1$. The arrows are the streamlines and color bars show the magnitude of the velocity field. (a) The viscosity ratio $\mu = 1$ and $\text{Ca} = 10^{-1}$ (surface tension is negligible). (b) $\mu = 0.1$ and $\text{Ca} = 10^{-1}$. The red line indicates the deformed interface. (c) $\mu = 0.1$ and $\text{Ca} = 10^{-6}$. We multiply $\zeta$ by 10 for a clearer view.}
\label{fig7}
\end{figure}
We further illustrate the effect of surface tension by plotting the flow field in Fig.~\ref{fig7}. For $\mu = 1$ and $\text{Ca} = 10^{-1}$, which is essentially a uniform fluid, the streamlines are symmetric about $x = 0$ [Fig.~\ref{fig7}(a)]. Replacing the substrate by a more viscous fluid with $\mu=0.1$ maintains this symmetric feature and makes the velocity decay faster along the $-y$ direction [Fig.~\ref{fig7}(b)]. The resulting deformation is also symmetric. The deformation profile is directly coupled to the $y$-component velocity, $w_1 = -U_{\mathrm{c}}\partial\zeta/\partial x$, i.e., the slop of $\zeta$ is opposite to the sign of $w_1$. Consistent with the prediction by Eq.~(\ref{eq51}), a large surface tension breaks the symmetry, leading to asymmetric interface deformation and streamlines [Fig.~\ref{fig7}(c)].   

The negative lift force observed at small Ca has a geometrical origin. In the elastic case, the interface is deformed by the pressure in such a way that the fluid flows from a narrower gap to a wider gap from the inlet to the outlet [Fig.~\ref{fig2}(b) inset], which can be approximated as an inclined slider with a positive slope. This simplified geometry resembles the thrust bearing that is well known to generate a positive lift force~\cite{Reynolds86, Hamrock04}. The viscous fluid layer at small Ca, on the other hand, creates an opposite geometry with a wider gap at the inlet and narrower gap at the outlet, which is approximated as a slider with a negative slope [Fig.~\ref{fig6}(b) inset] and generates a negative lift force. Define the gap profile as $H(x) = H_{\mathrm{c}}(1+b x/L)$, where $b$ is the slope. We solve the Reynolds equation within a finite domain $x\in [-L,L]$ with zero pressures at the boundaries. The resulting lift force $F/(p_0 L) = 12(\tanh^{-1}b-b)/b^2$, which is indeed positive for $b > 0$ and negative for $b < 0$.
\begin{figure}[t]
\centering
\includegraphics[bb=0 10 365 220, scale=0.66,draft=false]{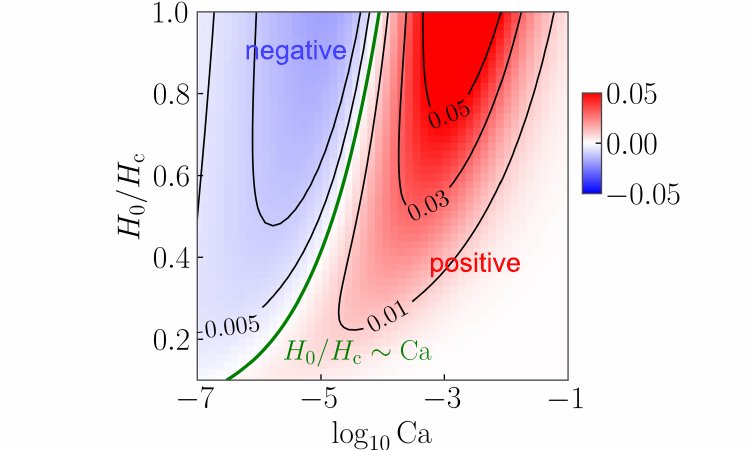}
\caption{The lift force as a function of $H_0/H_{\mathrm{c}}$ and $\mathrm{Ca}$. The green curve is the numerically best-fit boundary that separates the positive and negative regions.}
\label{fig8}
\end{figure}

Finally, we show the contour plot of $F$ versus $H_0$ and Ca (Fig.~\ref{fig8}). As $H_0$ is increased, the critical value of Ca that separates the positive and negative forces increases. The numerical fit of boundary is $H_0 \sim \mathrm{Ca}^{1.07}$, which is essentially a linear relation. Based on Eq.~(\ref{eq51}), this is a result of the balance between the viscous shear stress and the capillary restoring force, $\epsilon^3 (H_0/H_{\mathrm{c}})^3/\mathrm{Ca} \sim \mu (H_0/H_{\mathrm{c}})^2$, which yields $H_0/H_{\mathrm{c}} \sim \mu\epsilon^{-3} \mathrm{Ca}$. 

\section{Conclusions and discussion}\label{sec_4}
We have studied the lift force acting on a cylinder translating parallel to a viscoelastic fluid layer. Both the tangential and normal stresses are taken into account by the stress balance conditions at the layer surface. The surface slip velocity $U_{\mathrm{s}}$ is included explicitly for the first time in the calculation of the lift force. A wide range of layer thickness $H_0$ was explored ranging from thin ($H_0 \lesssim H_{\mathrm{c}}$) to thick ($H_0 \gg H_{\mathrm{c}}$) layer. 

We found that the lift force increases with the Deborah number De, matching the results of of the viscous fluid layer at small De and the elastic layer at large De. Combining numerical and asymptotic analysis, we showed that the tangential shear stress and the slip at the interface are important for the calculation of the lift force for $H_0 \lesssim H_{\mathrm{c}}$, giving important corrections to the previous analysis. We also found that at small De, where the layer elasticity is negligible, the lift force can become negative due to surface tension. We confirmed this result by directly solving Stokes equation for fluid layer and presented a qualitative explanation by drawing an analogy to the slider bearing with a negative slope.

The nonuniformity of the slip velocity $U_{\mathrm{s}}$ is the key to its contribution to the lift force, which suggests a way to affect particle motions by engineering the slippage of a nearby surface~\cite{Rinehart20}. \citet{Chandler20} also studied a similar lubrication problem, but they focused on a thick layer of $H_0 \gg H_{\mathrm{c}}$, and only considered the effect of normal stress. The effect of viscoelasticity on the lift force has been studied for a viscoelastic solid of infinite thickness~\cite{Pandey16}, which differs from the Maxwell model by an equilibrium modulus when $\omega \to 0$. Our result is consistent with Ref.~\cite{Pandey16} in that the lift force decreases as the layer becomes more viscous. The dependence on $H_0$ can be checked experimentally either by directly preparing substrates of different thickness~\cite{Saintyves16} or by changing the gap thickness $H_{\mathrm{c}}$~\cite{Zhang20}. Our results may also have implications for particle manipulations in microfluidics~\cite{Karimi13, Lu17, Gaetano17}, such as separation or focusing. We restrict ourselves to a steady translating motion of the cylinder. Controlled oscillatory motions may be employed as a contactless approach to probe the rheological properties of the substrate~\cite{Leroy11, Leroy12, Bertin21}.

\begin{acknowledgments}
S. Y. acknowledges support from National Natural Science Foundation of China (Grant No. 12247130). 
F. M. acknowledges supports from National Natural Science Foundation of China (Grant No. 12275332 and 12047503), Max Planck Society (Max Planck Partner Group) and Wenzhou Institute (Grant No. WIUCASQD2023009). 
Masao Doi acknowledges the startup fund of Wenzhou Institute, University of Chinese Academy of Sciences (WIUCASQD2022004) and Oujiang Laboratory (OJQDSP2022018). The computations of this work were conducted on the HPC cluster of ITP-CAS.

\end{acknowledgments}

\appendix
\section{Asymptotic solution of the viscous limit}\label{app_A}
Consider a thin viscous fluid layer with thickness $h(x)$ and viscosity $\eta$. The velocity in general has a quadratic form $v = A y^2 + B y +C$. The coefficients are determined by the definition $\bar{v} = 1/h\int_0^h v\,dy$ and the boundary conditions, $v|_{y = 0} = U_a$ and $v|_{y = h} = U_b$. Substituting into $\Phi = \eta/2\int_0^h (\partial_y v)^2\,dy$, we obtain the dissipation function written in terms of $\bar{v}$ for a single layer of viscous fluid~\cite{Man17},
\begin{equation}\label{eq_A1}
\Phi[U_a, U_b] = \frac{\eta}{h}\left[6\left(\bar{v}-\frac{U_a + U_b}{2}\right)^2 +\frac{(U_b-U_a)^2}{2}\right].
\end{equation}
It is convenient to work in the frame translating with the cylinder. For two layers of viscous fluids, the total dissipation function is the sum of the dissipation functions of the two layers, $\Phi = \Phi_1[-U_{\mathrm{c}}, U_{\mathrm{s}}] + \Phi_2[U_{\mathrm{s}}, 0]$. Minimizing $\Phi$ with respect to $U_{\mathrm{s}}$, we obtain $\Phi$~\cite{Hu22}, 
\begin{equation}\label{eq_A2}
\begin{aligned}
\Phi = \Bigg\{\frac{\eta_1\eta_2}{h_2\eta_1+h_1\eta_2}\Big[\frac{9 (\bar{v}_1-\bar{v}_2)^2}{2} + 3(\bar{v}_1-\bar{v}_2)U_{\mathrm{c}} \\
+ \frac{U_{\mathrm{c}}^2}{2} \Big] + \frac{3\eta_1}{2h_1} (\bar{v}_1+U_{\mathrm{c}})^2 + \frac{3\eta_2}{2h_2}\bar{v}_2^2\Bigg\}.
\end{aligned}
\end{equation}
The averaged velocities $\bar{v}_1$ and $\bar{v}_2$ are calculated by minimizing the Rayleighian [Eq.~(\ref{eq49})]. 

We define the following dimensionless variables, 
\begin{equation}\label{eq_A3}
\begin{aligned}
\hat{x} = \frac{x}{L},\ \hat{H}_0 = \frac{H_0}{H_{\mathrm{c}}},\ \hat{\zeta} = \frac{\zeta}{H_{\mathrm{c}}},\ \hat{H} = \frac{H}{H_{\mathrm{c}}}, \ \hat{p} = \frac{p}{p_0}.
\end{aligned}
\end{equation}
In steady state, $\partial_x (\bar{v}_1 h_1) = 0$. This leads to
\begin{equation}\label{eq_A4}
\frac{\epsilon^3 \hat{H}_0^3}{3\mathrm{Ca}}\frac{\partial^4 \hat{\zeta}}{\partial \hat{x}^4} - \frac{\partial \hat{\zeta}}{\partial \hat{x}} - \frac{1}{4}\mu \hat{H}_0^2 \frac{\partial}{\partial \hat{x}}\left[\hat{H}\frac{\partial \hat{p}}{\partial \hat{x}} - \frac{2}{\hat{H}}\right] = 0.
\end{equation}
The volume conservation condition, $\partial_x (\bar{v}_1 h_1 + \bar{v}_2 h_2) = 0$, leads to
\begin{equation}\label{eq_A5}
\begin{aligned}
\frac{\partial}{\partial \hat{x}}\left[\hat{H}^3 \frac{\partial \hat{p}}{\partial \hat{x}} + 6H + 6(1-\mu)\hat{\zeta}\left(1-\frac{\hat{H}^2}{2}\frac{\partial \hat{p}}{\partial \hat{x}}\right)\right]  \\
- 3 \frac{\epsilon^3 \hat{H}_0^2}{\mathrm{Ca}}\frac{\partial}{\partial \hat{x}}\left(\hat{H}\frac{\partial^3\hat{\zeta}}{\partial \hat{x}^3}\right) = 0.
\end{aligned}
\end{equation}
The zeroth-order equation is the same as Eq.~(\ref{eq33}). The first-order equations are
\begin{equation}\label{eq_A6}
\frac{\epsilon^3 \hat{H}_0^3}{3\mathrm{Ca}}\frac{\partial^4 \hat{\zeta}}{\partial \hat{x}^4} - \frac{\partial \hat{\zeta}}{\partial \hat{x}} - \frac{1}{4}\mu \hat{H}_0^2 \frac{\partial}{\partial \hat{x}}\left[\hat{H}\frac{\partial \hat{p}_2^{(0)}}{\partial \hat{x}} - \frac{2}{\hat{H}}\right] = 0,
\end{equation}
and 
\begin{equation}\label{eq_A7}
\begin{aligned}
\frac{\partial}{\partial \hat{x}}\left[\hat{H}^3 \frac{\partial \hat{p}^{(1)}}{\partial \hat{x}} + 6(1-\mu)\hat{\zeta}\left(1-\frac{\hat{H}^2}{2}\frac{\partial \hat{p}_2^{(0)}}{\partial \hat{x}}\right)\right] \\
- 3 \frac{\epsilon^3 \hat{H}_0^2}{\mathrm{Ca}}\frac{\partial}{\partial \hat{x}}\left(\hat{H}\frac{\partial^3\hat{\zeta}}{\partial \hat{x}^3}\right) = 0.
\end{aligned}
\end{equation}
Substituting $p_2^{(0)}$ into Eq.~(\ref{eq_A6}), we obtain
\begin{equation}\label{eq_A8}
\frac{\epsilon^3 \hat{H}_0^3}{3 \mathrm{Ca}}\frac{\partial^4 \hat{\zeta}}{\partial \hat{x}^4} - \frac{\partial \hat{\zeta}}{\partial \hat{x}} = 4\mu \hat{H}^2_0 \frac{\hat{x}(\hat{x}^2-1)}{(1+\hat{x}^2)^3},
\end{equation}
which is the non-dimensional form of Eq.~(\ref{eq51}).

\bibliographystyle{apsrev4-2}
\bibliography{main}

\end{document}